\documentclass[twocolumn]{physcon09} 
\usepackage[dcucite]{harvard}
\usepackage{epsfig}
\usepackage{amsfonts}
\usepackage{amsmath}
\usepackage{amstext}
\usepackage{amssymb}
\usepackage{graphicx}
\usepackage{color}

\title{Feedback-dependent control of stochastic synchronization in coupled neural systems}

\author{ \twlsfb Philipp H{\"o}vel, Sarang A. Shah, Markus A. Dahlem, and Eckehard Sch{\"o}ll$^*$
    \affiliation{
      Institut f{\"u}r Theoretische Physik, Technische Universit{\"a}t Berlin, Hardenbergstra{\ss}e 38, 10623
Berlin, Germany\\
      $^*$schoell@physik.tu-berlin.de
    }
}

\begin{document}



\maketitle

\begin{abstract}
We investigate the synchronization dynamics of two coupled noise-driven FitzHugh-Nagumo systems, 
representing two neural populations. For certain choices of the noise intensities and coupling 
strength, we find cooperative stochastic dynamics such as frequency synchronization and phase 
synchronization, where the degree of synchronization can be quantified by the ratio of
the interspike interval of the two excitable neural populations and the phase synchronization index, 
respectively. 
The stochastic synchronization can be either enhanced or suppressed by local time-delayed feedback 
control, depending upon the delay time and the coupling strength. 
The control depends crucially upon the coupling scheme of the control force,
i.e., whether the control force is generated from the activator or inhibitor signal, and applied to 
either component. For inhibitor self-coupling,  synchronization is
most strongly enhanced, whereas for activator self-coupling there exist distinct 
values of the delay time where the synchronization is strongly suppressed even in the strong 
synchronization regime. For cross-coupling strongly modulated  behavior is found.
\end{abstract}

\begin{keywords}
 Synchronization, noise, coupling, time-delayed feedback
\end{keywords}

\section{Introduction}
\label{sec:intro} 

The control of unstable or irregular states of nonlinear dynamic systems 
has many applications in different fields of physics, chemistry, biology, and medicine
\cite{SCH07}. 
A particularly simple and efficient control scheme is time-delayed  
feedback \cite{PYR92} which occurs naturally in a number of biological systems
including neural networks where both propagation delays and 
local neurovascular couplings lead to time delays \cite{HAK06,WIL99,GER02}. 
Moreover, time-delayed feedback loops might be deliberately implemented to
control neural disturbances, e.g., to suppress undesired synchrony 
of firing neurons in Parkinson's disease or epilepsy \cite{SCH94e,ROS04a,POP05}.
Here we study coupled neural systems subject to noise and time-delayed feedback 
\cite{HAU06,HOE09,SCH08,SCH09a}.
In particular we focus upon the question how stochastic synchronization of noise-induced
oscillations of two coupled neural populations can be controlled by time-delayed feedback,
and how robust this is with respect to different coupling schemes of the control force.

Time-delayed feedback control of noise-induced oscillations was
demonstrated in a single excitable system  \cite{JAN03,BAL04,PRA07,POT08}.
The simplest network configuration displaying features of neural interaction
consists of two coupled excitable systems. 

In order to grasp the complicated interaction between billions of neurons in large
neural networks, those are often lumped into groups of neural populations 
each of which can be represented as an effective excitable element that is mutually 
coupled to the other elements 
\cite{ROS04,POP05}. In this sense the simplest model which may reveal features 
of interacting neurons consists of two coupled neural oscillators. Each of these will be 
represented by a simplified FitzHugh-Nagumo (FHN) system \cite{FIT60,NAG62}, 
which is often used as a generic model for neurons, or more generally, excitable systems
\cite{LIN04}. 

This paper is organized as follow: We introduce the model equations and the feedback scheme in 
Sec.~\ref{sec:model}.
Sec.~\ref{sec:measures} is devoted to two measures of the stochastic synchronization.
These are investigated for different coupling schemes of the feedback in Sec.~\ref{sec:results}. 
Finally, we conlcude in Sec.~\ref{sec:conclusion}.

\section{Model Equations}
\label{sec:model} 
Neurons are excitable units which can emit spikes or bursts of
electrical signals, \textit{i.e.}, the system rests in a stable steady state,
but after it is excited beyond a threshold, it emits a pulse.  In the
following, we consider electrically coupled neurons
modelled by the FitzHugh-Nagumo system in the excitable regime:
\begin{subequations}\label{eq:neuro_2FHN} 
\begin{eqnarray}
  \label{eq:neuro_2FHN_1}
  \varepsilon_1 \frac{d u_1}{dt} &=& f\left(u_1,v_1\right) + C \left(u_2 -u_1\right)\\
  \frac{d v_1}{dt} &=& g\left(u_1,v_1\right) + D_1 \xi_1
\end{eqnarray}
\end{subequations}
\begin{subequations}\label{eq:neuro_2FHN_sys2} 
\begin{eqnarray}
  \label{eq:neuro_2FHN_2}
  \varepsilon_2 \frac{d u_2}{dt} &=& f\left(u_2,v_2\right) + C \left(u_1 -u_2\right)\\
  \frac{d v_2}{dt} &=& g\left(u_2,v_2\right) + D_2 \xi_2
\end{eqnarray}
\end{subequations}
with $f\left(u_i,v_i\right)=u_i-u_i^3/3-v_i$ and $g\left(u_i,v_i\right)=u_i+a$ ($i=1,2$).
The fast activator variables $u_i$ ($i = 1, 2$) refer to the transmembrane voltage,
and the slow inhibitor variables $v_i$ are related to the electrical conductance of the relevant ion 
currents. The parameter $a$ is the excitability
parameter. For the purposes of this paper, $a$ is fixed at $1.05$, such that there are no autonomous 
oscillations (excitable regime). $C$
is the diffusive coupling strength between $u_1$ and $u_2$. To introduce different time scales for 
both systems, $\varepsilon_1$ is set to $0.005$ and $\varepsilon_2$ is set to $0.1$. 
Both systems, when uncoupled, are driven entirely by
independent noise sources, which in the above equations are represented by $\xi_i$ ($i = 1, 2$, 
Gaussian white noise with zero mean and unity variance). $D_i$ is the noise intensity, and for the purposes of this
paper, $D_2$
will be held fixed at $0.09$ \cite{HAU06}.

\begin{figure}[t]
	\begin{center}
		\includegraphics[width=\linewidth]{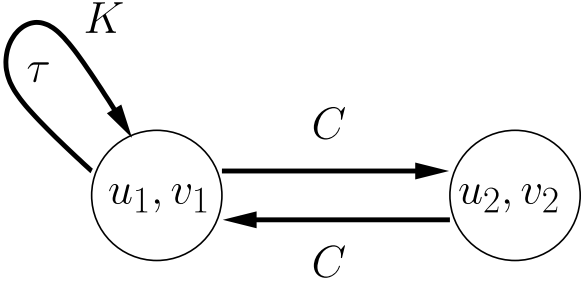}
	\end{center}
\caption{Schematic diagram of two coupled FitzHugh-Nagumo systems with time-delayed feedback applied to the first
subsystem. $K$ and $\tau$ denote the feedback gain and time delay, respectively, and $C$ is the coupling strength.}
\label{fig:neuro_schematic_2FHN_TDAS}
\end{figure}

The control force which we apply only to the first of the neural populations as schematically depicted in
Fig.~\ref{fig:neuro_schematic_2FHN_TDAS} is known as time-delay autosynchronization (TDAS) or time-delayed feedback
control. This method was initially introduced by Pyragas \cite{PYR92} for controlling periodic orbits in chaotic
systems. It has been effective in a variety of experimental applications 
at controlling oscillatory behavior and can be
easily implemented in many analog devices \cite{SCH07}. TDAS constructs a feedback $F$
from the difference between the current value of a control signal $w$ and the value for 
that quantity at time $t-\tau$.
The difference is then multiplied by the gain coefficient $K$
\begin{eqnarray}
 F(t) = K[w(t-\tau)-w(t)],
\end{eqnarray}
where $w$ determines which components of the system enter the feedback as will be discussed in the following.

The variable $w$ in the control force can be either the activator $u_1$ or the inhibitor $v_1$. Also, the
control force can either be applied to the activator or the inhibitor differential equation. These possibilities
lead to two self-coupling schemes ($uu$ and $vv$) where either the activator is coupled to the activator equation
or the inhibitor is coupled to the inhibitor equation, and two cross-coupling schemes ($uv$ and $vu$). 
Thus,
Eqs.~(\ref{eq:neuro_2FHN}) of the first subsystem can be rewritten including time-delayed feedback as
\begin{eqnarray}
  \label{eq:neuro_2FHN_control}
  \left(\begin{array}{c}
   \varepsilon_1 \frac{d u_1}{dt} \\
   \frac{d v_1}{dt}
  \end{array}\right)
  &=&
  \left(\begin{array}{c}
   f\left(u_1,v_1\right) + C \left(u_2 -u_1\right)\\
   g\left(u_1,v_1\right) + D_1 \xi_1
  \end{array}\right) \\
  &\,&+ K
  \left(\begin{array}{cc} 
    A_{uu} & A_{uv} \\
    A_{vu} & A_{vv}
  \end{array}\right)
  \left(\begin{array}{c}
    u_1(t-\tau)-u_1(t)\\
    v_1(t-\tau)-v_1(t)\\
  \end{array}\right),\nonumber
\end{eqnarray}
where the coupling matrix elements $A_{ij}$ with $i,j\in\{u,v\}$ define the specific coupling scheme.

Next, we will discuss cooperative stochastic dynamics resulting in frequency synchronization and
phase synchronization in the following Sections.

\section{Measures of Synchronization}
\label{sec:measures} 
A measure of frequency synchronization is the ratio of the interspike intervals (ISI) of 
the two neural populations \cite{HAU06,HOE09}. The respective average ISI of each neural population 
is denoted by $\langle T_1\rangle$ and $\langle T_2\rangle$. The ratio 
$\langle T_1\rangle/\langle T_2\rangle$ compares the
average time scales of both systems, where unity ratio describes two systems spiking at the same 
average frequency.
It is for this reason that the ISI ratio is often considered as a measure of frequency synchronization. 
It does not contain information about the phase of synchronization, and a given ISI ratio can also 
result from different ISI distributions.

\begin{figure}[t]
	\begin{center}
		\includegraphics[width=\linewidth]{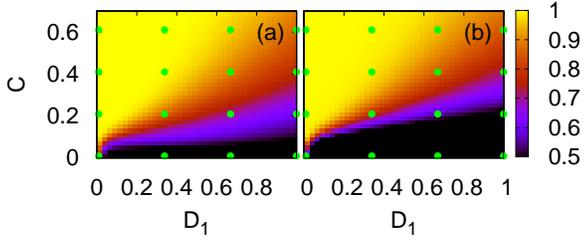}
	\end{center}
\caption{(Color online) Panels (a) and (b) show the ratio of interspike intervals $\langle T_1\rangle /\langle
T_2\rangle
$ and the phase synchronization index $\gamma$ of the two subsystems as color code in dependence on the coupling
strength $C$ and noise intensity $D_1$, respectively. No control is applied to the system. The dots mark the parameter
choice for different synchronization regimes used in the following. Other parameters: $\varepsilon_1=0.005$,
$\varepsilon_2=0.1$, $a=1.05$, and $D_2=0.09$.}
\label{fig:HOE09-figure1_mean_ISI_phase_sync}
\end{figure}

In order to account for the phase difference between two systems, one can define a  
phase 
\cite{PIK96,PIK01,HAU06}
\begin{eqnarray}\label{eq:neuro_phasediff} 
   \varphi(t) &=& 2\pi\frac{t-t_{i-1}}{t_{i}-t_{i-1}}+2\pi (i-1)
\end{eqnarray}
where $i = 1, 2,\dots$. $t_i$ denotes the time of the $i$th spike. The phase difference between two consecutive spikes
is
$2\pi$. The phase difference of 1:1 synchronization is 
\begin{eqnarray}
 \Delta\varphi(t) &=& \left|\varphi_{1}(t) - \varphi_{2}(t)\right|,
\end{eqnarray}
where $\varphi_1(t)$ and $\varphi_2(t)$ are the phases of the first and second system, respectively.  
Two systems that are phase synchronized at a given time satisfy $\Delta\varphi=0$. 
Finally, the overall time-averaged phase synchronization of two systems can be quantified using the 
synchronization index
\begin{eqnarray} 
 \label{eq:neurp_gamma} 
 \gamma = \sqrt{\langle\cos \Delta \varphi(t)\rangle^2+\langle\sin \Delta \varphi(t)\rangle^2}.
\end{eqnarray}
A value of $0$ indicates no synchronization, while a value of unity indicates perfect synchronization.

Figure~\ref{fig:HOE09-figure1_mean_ISI_phase_sync} depicts both measures for stochastic 
synchronization in the $\left(D_1,C\right)$ plane, both exhibiting very similar behavior. 
Panel (a) refers to the frequency synchronization characterized by the ratio of the average
ISIs $\langle T_1\rangle/\langle T_2\rangle$ and panel (b) shows the phase synchronization index $\gamma$. The green
dots mark parameter values used in Sec.~\ref{sec:results}. Note that both panels share the same color 
code. For a small
value of $D_1$ and large coupling strength, the two subsystems display well synchronized behavior, 
$\langle
T_1\rangle/\langle T_2\rangle\approx 1$ and $\gamma\approx 1$. The timescales in the interacting 
systems adjust themselves to
$1:1$ synchronization. On average, they show the same number of spikes and the two subsystems are in-phase which is
indicated by yellow color. The two subsystems are less synchronized in the dark blue and black 
regions. 

In the following we show the ratio of the average interspike interval $\langle T_1 \rangle / \langle T_2
\rangle$ and the phase synchronization index $\gamma$ which are color coded in the $(\tau,K)$ plane
for fixed combinations of $D_1$ and $C$. For each coupling scheme of time-delayed feedback control 
(cross-coupling schemes $uv$ and $vu$ and self-coupling schemes $uu$ and $vv$) we present a selection 
of $\left(D_1,C\right)$ values. In all cases, only one element of the 
coupling matrix ${\bf A}$ is equal to unity and all other elements are zero.

\section{Coupling Schemes}
\label{sec:results}
After the introduction of the system and the coupling schemes, we will present results on frequency and phase
synchronization in the following. We consider $16$ different combinations of the noise intensity  $D_1$ and the coupling
strength $C$ which are marked as green dots in  Fig.~\ref{fig:HOE09-figure1_mean_ISI_phase_sync}. The ordering of panels
in Figs.~\ref{fig:ratio_uu} to \ref{fig:gamma_vv} is the following: The rows correspond to fixed coupling strength
chosen as $C=0.01,0.21,0.41$, and $0.61$ from bottom to top. The columns in each figure are calculated for constant
noise intensity $D_1=0.01,0.34,0.67$, and $1.0$ from left to right.

\subsection{Frequency Synchronization}
\label{subsec:ratio}
\begin{figure}[h!]
	\begin{center}
		\includegraphics[width=\linewidth]{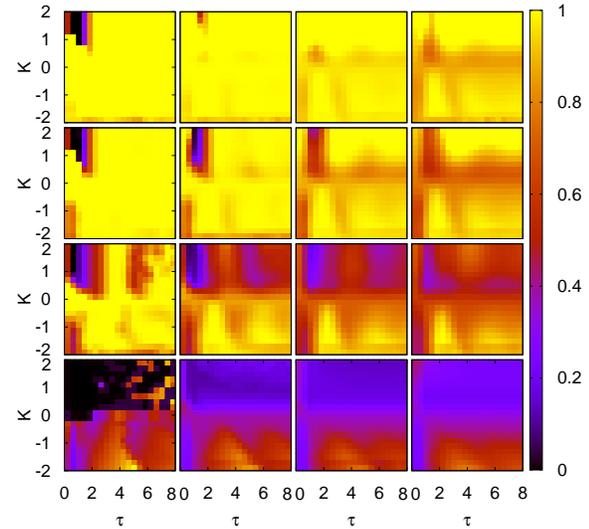}
	\end{center}
\caption{$uu$-coupling: Ratio of average interspike intervals 
$\langle T_1 \rangle / \langle T_2 \rangle$ as color code. Rows and columns correspond to constant 
coupling strength $C$  and noise intensity $D_1$, respectively, 
as marked in Fig.~\ref{fig:HOE09-figure1_mean_ISI_phase_sync} as green dots, and specified in the text. 
Other parameters as in Fig.~\ref{fig:HOE09-figure1_mean_ISI_phase_sync}.}
\label{fig:ratio_uu}
\end{figure}

\begin{figure}[h!]
	\begin{center}
		\includegraphics[width=\linewidth]{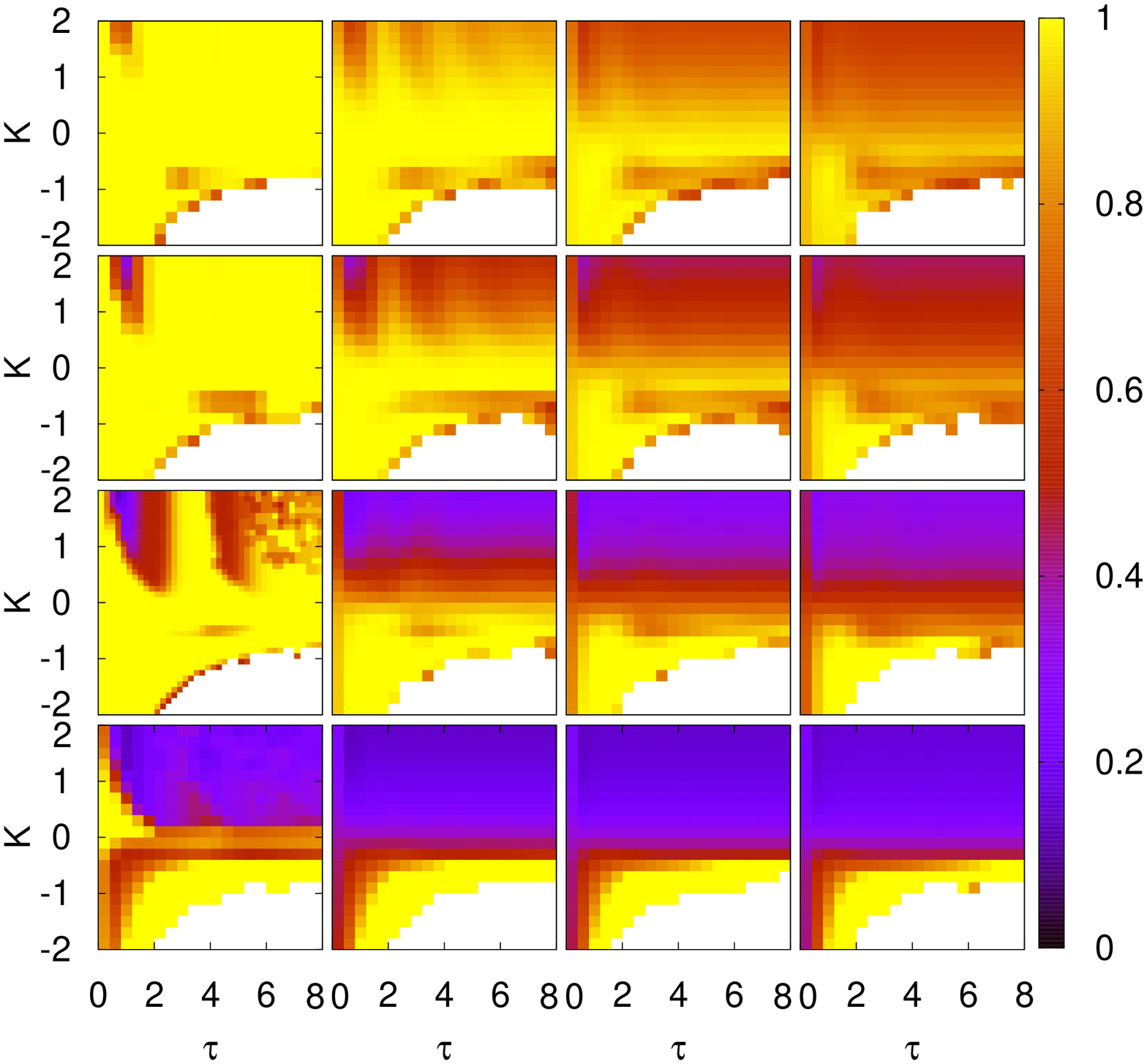}
	\end{center}
\caption{$uv$-coupling: Ratio of average interspike intervals $\langle T_1 \rangle / \langle T_2 \rangle$ as color code. Rows and columns correspond to constant coupling strength $C$  and noise intensity $D_1$, respectively. Other parameters as in
Fig.~\ref{fig:HOE09-figure1_mean_ISI_phase_sync}.}
\label{fig:ratio_uv}
\end{figure}

\begin{figure}[h!]
	\begin{center}
		\includegraphics[width=\linewidth]{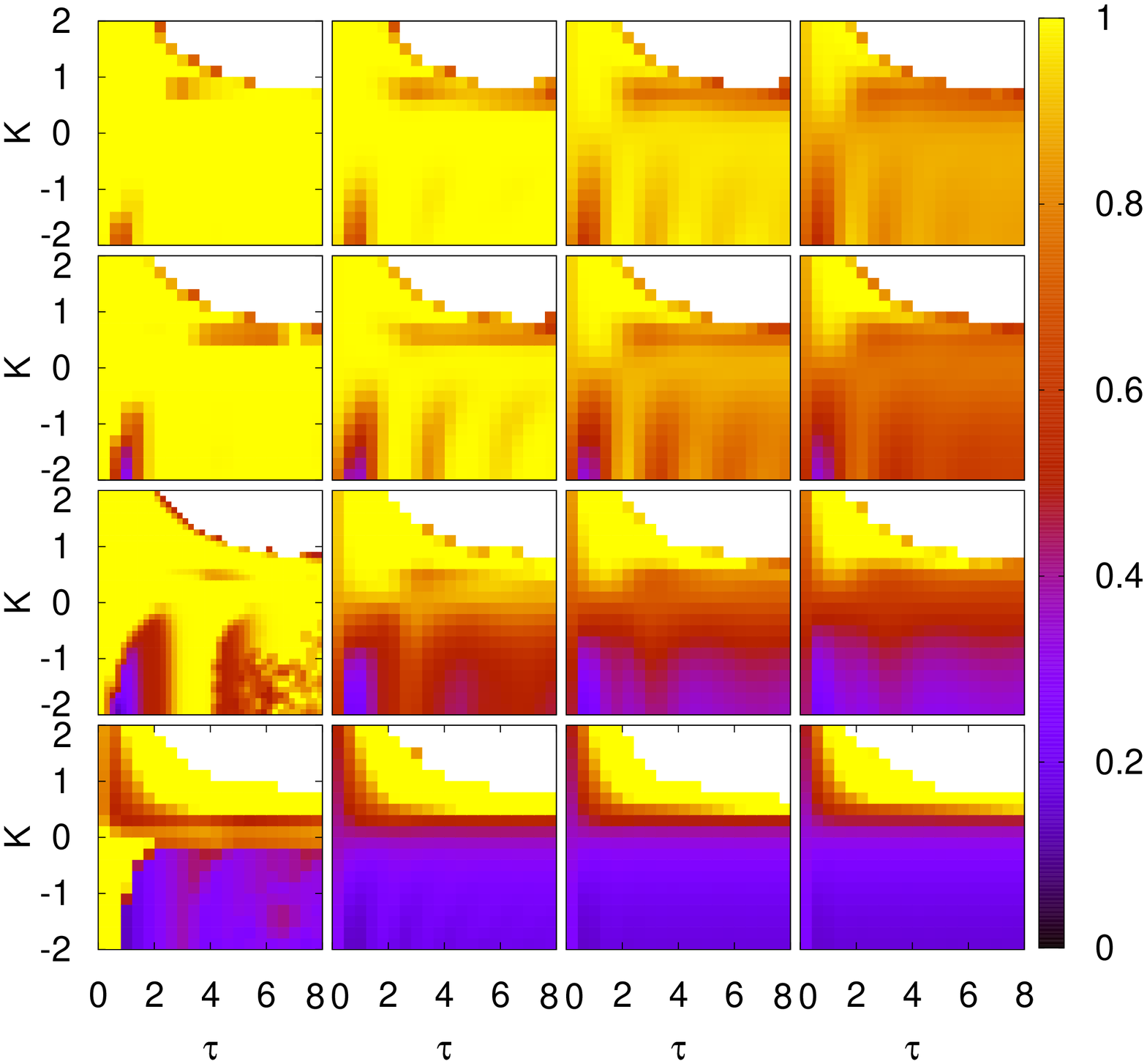}
	\end{center}
\caption{$vu$-coupling: Ratio of average interspike intervals $\langle T_1 \rangle / \langle T_2 \rangle$ as color code. Rows and columns correspond to constant coupling strength $C$  and noise intensity $D_1$, respectively. Other parameters as in
Fig.~\ref{fig:HOE09-figure1_mean_ISI_phase_sync}.}
\label{fig:ratio_vu}
\end{figure}

\begin{figure}[h!]
	\includegraphics[width=\linewidth]{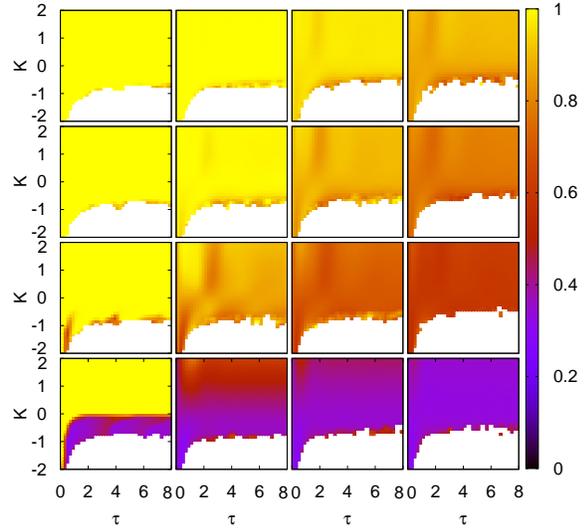}
\caption{$vv$-coupling: Ratio of average interspike intervals $\langle T_1 \rangle / \langle T_2 \rangle$ as color code. Rows and columns correspond to constant coupling strength $C$  and noise intensity $D_1$, respectively. Other parameters as in
Fig.~\ref{fig:HOE09-figure1_mean_ISI_phase_sync}.}
\label{fig:ratio_vv}
\end{figure}

Figures~\ref{fig:ratio_uu} to \ref{fig:ratio_vv} show frequency synchronization measured by the ratio of average
interspike intervals $\langle T_1 \rangle / \langle T_2 \rangle$ calculated from the summarized activator variable
$u_{\Sigma}=u_1+u_2$ as color code in dependence on the feedback gain $K$ and the time delay $\tau$. The system's
parameters are fixed in each panel as described above. Figures~\ref{fig:ratio_uu} and \ref{fig:ratio_vv} correspond to
self-coupling ($uu$- and $vv$-coupling) and Figs.~\ref{fig:ratio_uv} and \ref{fig:ratio_vu} depict the  cross-coupling
schemes ($uv$- and $vu$-coupling). The dynamics in the white regions is outside the excitable regime and does not show
noise-induced spiking, but rather the system exhibits large-amplitude self-sustained oscillations.

One can see that appropriate tuning of the control parameters leads to enhanced or deteriorated synchronization
displayed by bright yellow and dark blue areas, respectively. In each figure, all panels show 
qualitatively similar features like a modulation of the ratio $\langle T_1 \rangle / \langle T_2 \rangle$ 
whose range between maximum and
minimum depends on $D_1$ and $C$. Comparing the rows, the systems are less (more strongly) 
synchronized for small (large) 
values of $C$ indicated by dark blue (yellow) color. As the noise intensity $D_1$ increases, the 
dynamics of the coupled
subsystems is more and more noise-dominated and the dependence on the time delay $\tau$ becomes less 
pronounced.

Note the symmetry in the cross-coupling schemes shown as Figs.~\ref{fig:ratio_uv} and 
\ref{fig:ratio_vu} between $K$
and its negative value $-K$ for the inverse cross-coupling. The reason is that enhancing the 
activator yields a similar effects on the dynamics as diminishing the inhibitor variable. 

\subsection{Phase Synchronization}
\label{subsec:gamma}

\begin{figure}[h!]
	\begin{center}
		\includegraphics[width=\linewidth]{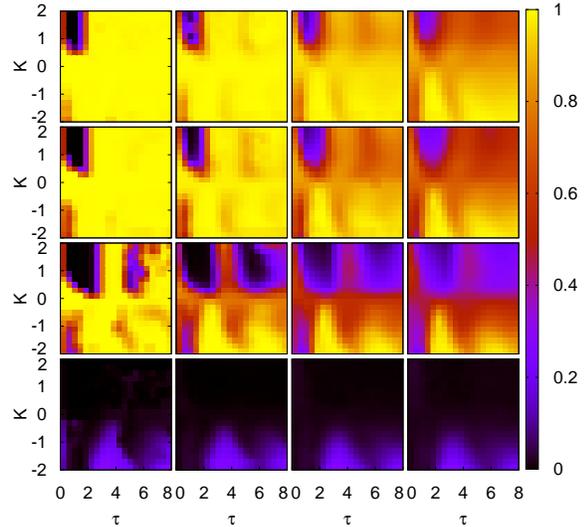}
	\end{center}
\caption{$uu$-coupling: Phase synchronization index $\gamma$. Noise intensity $D_1$ and coupling strength $C$ chosen as described in Sec.~\ref{subsec:ratio}. Other parameters as in
Fig.~\ref{fig:HOE09-figure1_mean_ISI_phase_sync}.}
\label{fig:gamma_uu}
\end{figure}

\begin{figure}[ht]
	\begin{center}
		\includegraphics[width=\linewidth]{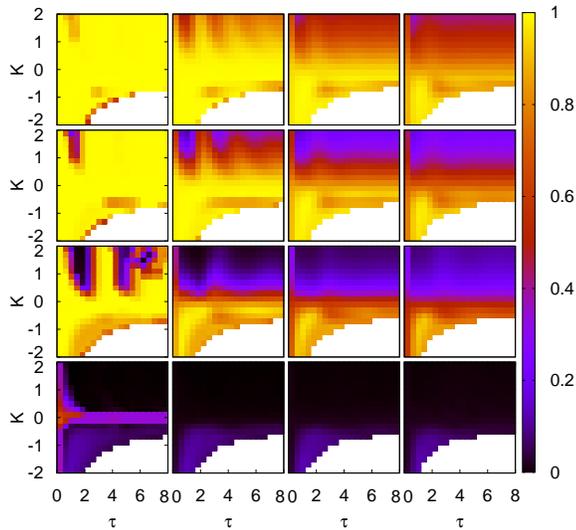}
	\end{center}
\caption{$uv$-coupling: Phase synchronization index $\gamma$. Noise intensity $D_1$ and coupling strength $C$ chosen as described in Sec.~\ref{subsec:ratio}. Other parameters as in Fig.~\ref{fig:HOE09-figure1_mean_ISI_phase_sync}.}
\label{fig:gamma_uv}
\end{figure}

\begin{figure}[ht]
	\begin{center}
		\includegraphics[width=\linewidth]{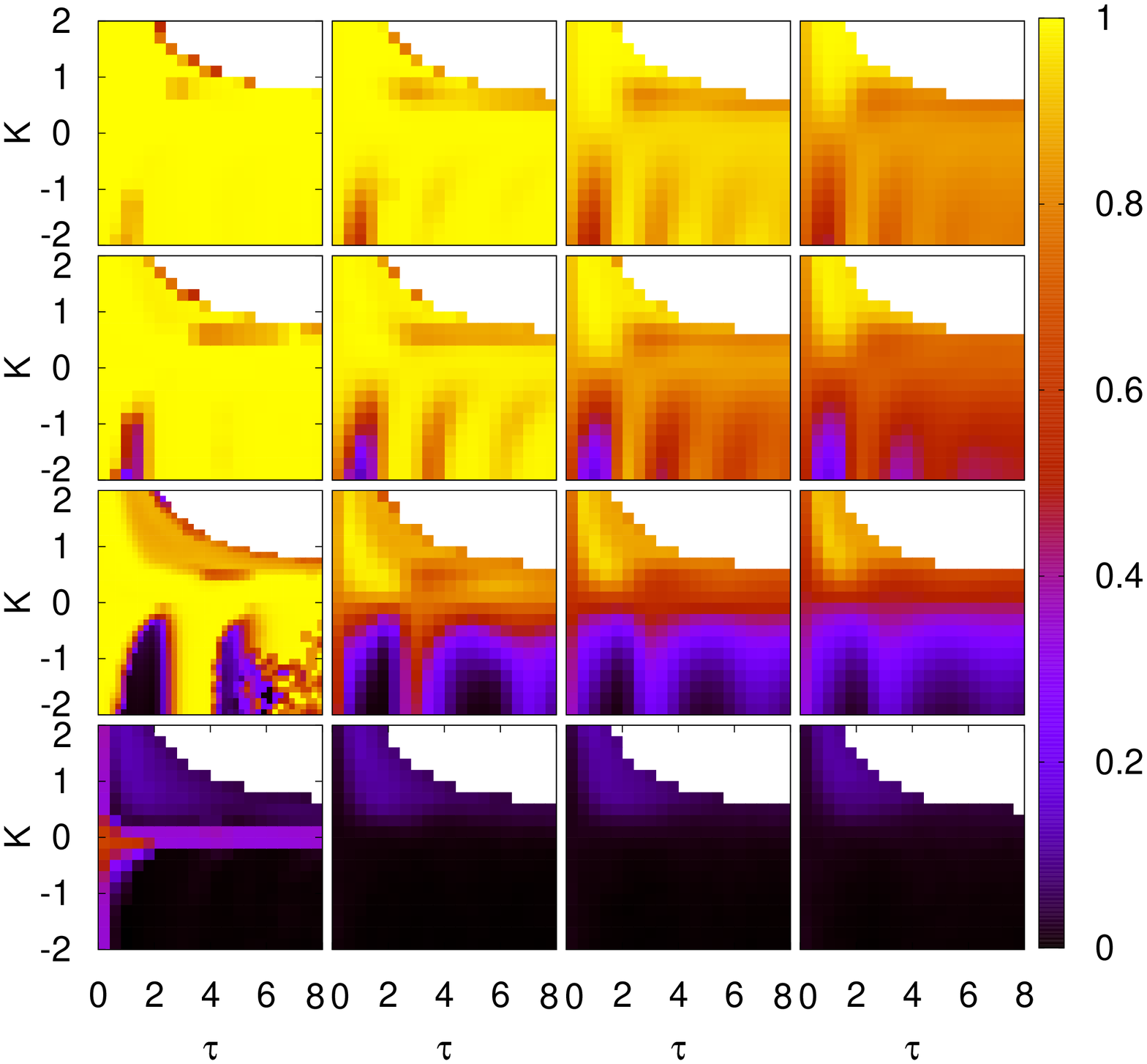}
	\end{center}
\caption{$vu$-coupling: Phase synchronization index $\gamma$. Noise intensity $D_1$ and coupling strength $C$ chosen as described in Sec.~\ref{subsec:ratio}. Other parameters as in Fig.~\ref{fig:HOE09-figure1_mean_ISI_phase_sync}.}
\label{fig:gamma_vu}
\end{figure}

\begin{figure}[ht]
	\includegraphics[width=\linewidth]{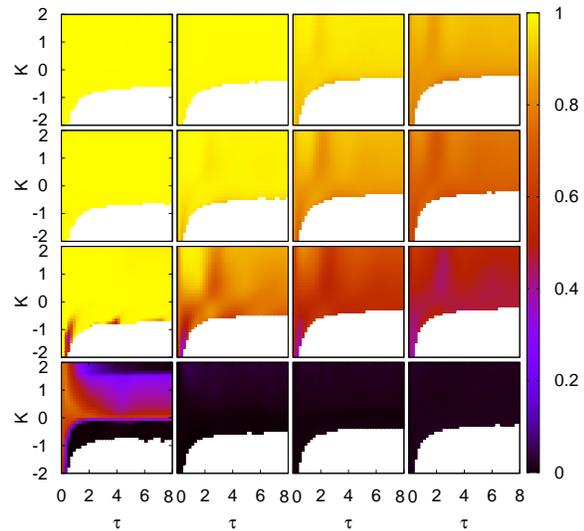}
\caption{$vv$-coupling: Phase synchronization index $\gamma$. Noise intensity $D_1$ and coupling strength $C$ chosen as
described in Sec.~\ref{subsec:ratio}. Other parameters as in Fig.~\ref{fig:HOE09-figure1_mean_ISI_phase_sync}.}
\label{fig:gamma_vv}
\end{figure}

Figures~\ref{fig:gamma_uu} to \ref{fig:gamma_vv} depict the phase synchronization index $\gamma$ as color code depending
on the control parameters $K$ and $\tau$ for $uu$-, $uv$-, $vu$-, and $vv$-coupling, respectively. The noise intensity
$D_1$ and coupling strength $C$ are fixed for each panel as described in Sec.~\ref{subsec:ratio}.

Comparing Figs.~\ref{fig:gamma_uu} to \ref{fig:gamma_vv} with the respective plots for frequency 
synchronization, i.e., Figs.~\ref{fig:ratio_uu} to \ref{fig:ratio_vv}, one can see that both types of 
synchronization coincide qualitatively, but the phase synchronization index is more sensitive to
the modulation features. Similar to the case of frequency synchronization, time delayed feedback can 
lead to either enhancement or suppression of phase synchronization depending on the specific choice 
of the feedback gain $K$ and time delay $\tau$ indicated by yellow and dark blue regions. 
In general, these effects become less sensitive on the time delay as $D_1$ increases. For larger 
values of $C$, the two subsystems show enhanced phase synchronization.

\section{Conclusion}
\label{sec:conclusion} 
In summary, we have shown that stochastic synchronization in two coupled neural populations can be
tuned by local time-delayed feedback control of one population. Synchronization can be either enhanced 
or suppressed, depending upon the delay time and the coupling strength. The control dependents crucially
upon the coupling scheme of the control force. For inhibitor self-coupling ($vv$) synchronization is
most strongly enhanced, whereas for activator self-coupling ($uu$) there exist distinct 
values of $\tau$ where the synchronization is strongly suppressed even in the strong synchronization
regime. For cross-coupling ($uv$, $vu$) there is mixed behavior, and both schemes exhibit a strong
symmetry with respect to inverting the sign of $K$. These observations might be important in the
context of the deliberate application of control with the aim of suppressing synchronization, e.g. 
as therapeutic measures for Parkinson's disease.

\section*{Acknowledgements}
This work was supported by DFG in the framework of Sfb 555 (Complex Nonlinear Processes). S.~A.~S. acknowledges
support of the Deutsche Akademische Austausch\-dienst (DAAD) in the framework of the program Research Internships in
Science and Engineering (RISE).


\end{document}